\documentclass[conference]{IEEEtran}
\IEEEoverridecommandlockouts

\usepackage{graphicx}
\usepackage{eurosym}
\usepackage{array}
\usepackage{subfigure}
\usepackage{amssymb,amsmath,amsfonts}
\usepackage{booktabs}
\usepackage{multirow}
\usepackage[table]{xcolor}
\usepackage{arydshln}
\usepackage{bbm}
\usepackage{cite}

\usepackage{algorithm}
\usepackage[noend]{algpseudocode}

\DeclareMathAlphabet{\mathcal}{OMS}{cmsy}{m}{n}

\algdef{SE}[DOWHILE]{Do}{doWhile}{\algorithmicdo}[1]{\algorithmicwhile\ #1}%

\usepackage{float}
\usepackage{psfrag}
\usepackage{hyperref}
\hypersetup{
    colorlinks,
    citecolor=blue,
    filecolor=blue,
    linkcolor=blue,
    urlcolor=blue,
    pdfstartview={XYZ null null 1.00}
}
\definecolor{gray1}{gray}{0.7}
\definecolor{gray2}{gray}{0.9}

\renewcommand{\deg}{$^{\circ}$}
\newcommand{\decBayes}{\textit{Bayes-Swarm}}
\newcommand{\alphaCoef}{exploitation coefficient }
\newcommand{\bx}{\mathbf{x}}

\begin{document}

\title{Decentralized Informative Path Planning with Exploration-Exploitation Balance for Swarm Robotic Search
\thanks{Copyright\copyright 2019 ASME. Personal use of this material is permitted. Permission from ASME must be obtained for all other uses, in any current or future media, including reprinting/republishing this material for advertising or promotional purposes, creating new collective works, for resale or redistribution to servers or lists, or reuse of any copyrighted component of this work in other works.}
}

\author{\IEEEauthorblockN{Payam Ghassemi\IEEEauthorrefmark{1}, and
Souma Chowdhury\IEEEauthorrefmark{2}}
\IEEEauthorblockA{\textit{Department of Mechanical and Aerospace Engineering} \\
\textit{University at Buffalo}\\
Buffalo, NY, 14260\\
Email: \IEEEauthorrefmark{1}payamgha@buffalo.edu,
\IEEEauthorrefmark{2}soumacho@buffalo.edu}}

\maketitle

\begin{abstract}
 \emph{Swarm robotic search is concerned with searching targets in unknown environments (e.g., for search and rescue or hazard localization), using a large number of collaborating simple mobile robots. In such applications, decentralized swarm systems are touted for their task/coverage scalability, time efficiency, and fault tolerance. To guide the behavior of such swarm systems, two broad classes of approaches are available, namely nature-inspired swarm heuristics and multi-robotic search methods. However, simultaneously offering computationally-efficient scalability and fundamental insights into the exhibited behavior (instead of a black-box behavior model), remains challenging under either of these two class of approaches. In this paper, we develop an important extension of the batch Bayesian search method for application to embodied swarm systems, searching in a physical 2D space. Key contributions lie in: 1) designing an acquisition function that not only balances exploration and exploitation across the swarm, but also allows modeling knowledge extraction over trajectories; and 2) developing its distributed implementation to allow asynchronous task inference and path planning by the swarm robots. The resulting collective informative path planning approach is tested on target search case studies of varying complexity, where the target produces a spatially varying (measurable) signal. Significantly superior performance, in terms of mission completion efficiency, is observed compared to exhaustive search and random walk baselines, along with favorable performance scalability with increasing swarm size.}
\end{abstract}


\section{INTRODUCTION} \label{sec:intro}
Swarm robotic search is concerned with searching and/or localizing targets in unknown environments with a large number of collaborative robots. Potential applications include source
localization of gas leakage~\cite{baetz2009mobile}, nuclear meltdown tracking~\cite{nagatani2013emergency},  chemical plume tracing~\cite{li2006moth}, radio
source localization~\cite{song2012simultaneous}, cooperative
foraging~\cite{sugawara2004foraging}, and oil spill mapping~\cite{senga2007development, odonkor2019distributed}.
Swarm robotic systems demonstrate mission efficiency, fault tolerance, and scalable coverage advantages~\cite{desilva2012development,ghassemi2018decentralized} compared to sophisticated standalone systems. In swarm robotic search, a major challenge is in designing computationally lightweight algorithms that allow effective task-planning within the swarm of robots~\cite{pugh2007inspiring}, one that maximizes search efficiency and mitigates conflicts. 
In this paper, we consider searching targets that emit a spatially varying signal (aka. a source localization problem) using a swarm of robots in 2D space. The online planning problem solved by the robots is then posed as finding a set of waypoints that maximize some measure of collective search efficiency~\cite{kalmar2017multiagent}. For this purpose, we formulate, implement and test a novel decentralized algorithm, founded on the batch Bayesian search formalism, that not only tackles the balance between exploration and exploitation, but also allows asynchronous decision-making within the swarm.
The proposed formulation is tested over a set of case studies involving varying number of robots and target sources. The remainder of this section briefly surveys the literature on swarm robotic search, and states the objectives of this paper.

\subsection{Single-robot vs. Multi-robot Search}
%
Various search strategies for single-robot system have been surveyed in~\cite{benkoski1991survey}. Most of these works are theoretical in nature and applicable to a single robot searching for single, multiple, static or dynamic targets. However, in time-sensitive applications involving large areas and multiple signal sources~\cite{tan2013research}, a team of robots can broaden the scope of operational capabilities through distributed remote sensing, scalability and parallelism (both in terms of task execution and information gathering)~\cite{odonkor2019distributed}. The multi-robot search paradigm (typically involving 10 or less number of robots) uses concepts such as predefined lanes or patterns~\cite{beard2003multiple}, space filling curves,~\cite{flint2003cooperative}, Voronoi-based methods~\cite{vincent2004framework, cortes2004coverage}, control theory, team theory~\cite{rajnarayan2003multiple}, and uncertainty reduction methods \cite{sujit2009negotiation}. 
Among these approaches, the ones suited for search in unknown unstructured environments generally do not scale well from the multi-robotic to the swarm-robotic paradigm (where the latter involves 10's to 100's of robots). 

\subsection{Swarm Robotic Search}
The class of approaches more popular in guiding the agents' behavior in scalable swarm robotic search is often based on nature-inspired swarm intelligence (SI) algorithms ~\cite{kennedy2010particle,krishnanand2009glowworm,senanayake2016search,dhameliya2018prototyping}. Variations of these algorithms are otherwise also used for performing optimization on highly nonconvex multimodal functions \cite{chowdhury2013mixed}, which can be perceived as non-embodied search in an $n$-dimensional space. A few examples of nature-inspired swarm robotic search is given here. Pugh and Martinoli~\cite{pugh2007inspiring} proposed an algorithm based on particle swarm concepts for swarm robotic search considering a stationary single target case. Jatmiko et al.~\cite{jatmiko2006pso} studied a particle swarm-based algorithm for odor source localization using a team of 20 robots. 
They showed that effectiveness of these methods rely on adaptive parameters (e.g., decreasing the inertia weight linearly during the search) to be successful to find the source location. A comprehensive (albeit bit dated) review of work in swarm robotic search (including nature inspired methods) can be found in \cite{brambilla2013swarm}.

\textit{Multi-modal search environments}: The use of robotic systems to locate a single gradient source has been investigated in the literature~\cite{mcgill2009comparing}. However, the localization of multiple gradient sources or the maximum strength source in the presence of other weaker sources (i.e., a multi-modal spatial distribution of signal strength) has received much less attention. A notable exception is the gradient search based distributed algorithm reported by Krishnanand et al.~\cite{krishnanand2006glowworm}. 
McGill and Taylor~\cite{mcgill2009comparing} however showed that the former approach~\cite{krishnanand2006glowworm} is not able to locate all sources if the initial robot distribution does not cover the search area (an impractical assumption).

In adopting population-based search algorithms, typically used in optimization (a non-embodied process), for physical swarm-robotic search, we need to appreciate two main differences in the nature of the sampling process. We see the process of deciding the next point to be explored by each agent as ``sampling" -- which is system evaluation in optimization and signal measurement in robotic search. With that perspective, the main differences are in: (1) \textit{sampling cost:} unlike in optimization, each sample may require a different energy/time investment by robots depending upon the distance of the next waypoint and the operating environment; and (2) \textit{sampling over path:} robots are able to gather samples (signal measurements) over their path, unlike in optimization where each population member evaluates the system only at their next point. Thus, knowledge generation in robotic search occurs over trajectories, while in optimization it occurs at discrete points in the search space. Although, the sampling cost aspect has been considered in the aforementioned swarm intelligence-based methods, the ``sampling over path" factor has received minimal attention. In our proposed approach, we adopt the Bayesian optimization and informative path planning principles for swarm robotic search. Here, robots employ informative path planning to generate paths that simultaneously balances reduction of the knowledge uncertainty and getting closer to the global source. This approach is such designed as to allow collectively assisting in reducing uncertainty, but not necessarily requiring all robots to converge on the source. Given that robotic search is in general restricted to 2D or 3D space (higher dimensional state spaces are possible), a new batch Bayesian approach is expected to be both effective and computationally frugal for this purpose.  

Moreover, with swarm-intelligence based methods, the dependence of the (at times astonishingly competitive) \textit{emergent} behavior on heuristics raises questions of dependability and explainability (a particular concern in applications requiring human-swarm teaming~\cite{kolling2016human}). 
Now, the search problem can be thought of as comprising two main steps: task perception (identifying/updating the signal spatial model) and task selection (waypoint planning). In swarm intelligence methods, the two steps are not separable, and a spatial model is not explicit. In our proposed approach, the processes are inherently decoupled -- the robots exploit Gaussian Processes (GP) to model the signal distribution knowledge (task inference stage) and solves a 2D optimization over the acquisition function to decide waypoints (task selection). Such an approach is expected to provide greater explainability and ease of debugging any performance shortfalls. 

\subsection{Objective of This Paper}
%
The primary objective of this paper is to develop (an explainable) decentralized and asynchronous swarm robotic search algorithm, subject to the following assumptions: 
i) all robots are equipped with precise localization; and ii) each robot can communicate their knowledge, state and decisions with all neighbors (full observability) at waypoints. Within this context, the primary contributions of this paper lies in the following developments: \textit{1) a novel decentralized algorithm (\decBayes) that extends Gaussian process modeling (to update over trajectories) and integrates physical robot constraints and other robots' decisions to perform informative path planning -- simultaneously mitigating knowledge uncertainty and getting closer to the source; and 2) a simulated parallelized implementation of {\decBayes} to allow asynchronous search planning over complex multi-modal signal distributions.} The performance of {\decBayes} is also compared with that of a random-walk baseline.

The remaining portion of the paper is organized as follows: The next section presents the problem definition and GP modeling. Then our proposed decentralized algorithm (\decBayes) is described. Numerical experiments and results, encapsulating the performance of these methods on different-sized swarm and a parametric analysis of the proposed decentralized method are then presented. The paper ends with concluding remarks.

\section{BACKGROUND}
%
\subsection{Gaussian Process Model}
Gaussian process (GP) models~\cite{rasmussen2003gaussian} are probabilistic surrogate models that have been used successfully in different applications such as modeling the objective function in Bayesian optimization~\cite{snoek2012practical}. If we have a set of $n$ observations of an environment, $\mathcal{D}={\mathbf{x}_i,y_i|i=1\dots n}$, then we can write the following equation by assuming that the observed values $y$ differ from the function $f(\mathbf{x})$ values by an additive noise $\epsilon$, where $\mathbf{x}$ denotes an input vector:
\begin{equation}
y = f(\mathbf{x}) + \epsilon    
\end{equation}
By assuming the noise follows an independent, identically distributed Gaussian distribution with zero mean and variance $\sigma^2_n$, we have $\epsilon \sim \mathcal{N}(0,\sigma_n^2)$.
The function $f(\mathbf{x})$ can be estimated by a GP with mean function $\mu(\bx)$ and covariance kernel $\sigma^2(\bx)$ given by: 
\begin{equation}
    f(\mathbf{x}) \sim \mathcal{GP}\big(\mu(\mathbf{X}),\sigma^2(\mathbf{X})\big)
\end{equation}
where,
\begin{align}
\mu(\mathbf{x}) &= \Lambda(\mathbf{x})(\mathbf{y}-\Phi\beta)\\
\sigma^2(\mathbf{x}) &= k(\mathbf{x},\mathbf{x}) - \Lambda(\mathbf{x})\mathbf{k}_n(\mathbf{x})\\
\Lambda(\mathbf{x}) &= \mathbf{k}_n(\mathbf{x})^T [\mathbf{K} +\sigma_n^2(\mathbf{x})\mathbf{I}]^{-1}
\end{align}
Here $\Phi$ is the vector of explicit basis functions and $\mathbf{K} = \mathbf{K}(\mathbf{X},\mathbf{X}|\theta)$ is the covariance function matrix such that $(\mathbf{K})_{ij} = k(\mathbf{x}_i, \mathbf{x}_j)$, and $\mathbf{k}_n(\mathbf{x}) = [k(\mathbf{x}_1,\mathbf{x}), \dots, k(\mathbf{x}_n,\mathbf{x})]^T$.
In this paper, the hyper-parameters of the GP model are optimized by maximizing the log-likelihood $P$ as a function of $\beta,\theta,\sigma_f^2$:
\begin{equation}
\hat{\beta},\,\hat{\theta},\,\hat{\sigma}^2 = \text{arg}\max_{\beta,\theta,\sigma^2}\log{P(\mathbf{y}|\mathbf{X},\beta,\theta,\sigma_f^2)}
\end{equation}
where,
\begin{equation}
\begin{aligned}
\log{P(\mathbf{y}|\mathbf{X},\beta,\theta,\sigma^2)} = &-\frac{1}{2}(\mathbf{y}-\Phi\beta)^T\Lambda(\mathbf{x})^{-1}(\mathbf{y}-\Phi\beta)\\
&- \frac{N_s}{2}\log{2\pi}-\frac{1}{2}\log{|\Lambda(\mathbf{x})|}
\end{aligned}
\end{equation}

\section{SWARM BAYESIAN ALGORITHM}\label{sec:decBayes}
\subsection{Bayes-Swarm: Overview}
\begin{figure}[!hpt]
\centering
\includegraphics[width=0.45\textwidth]{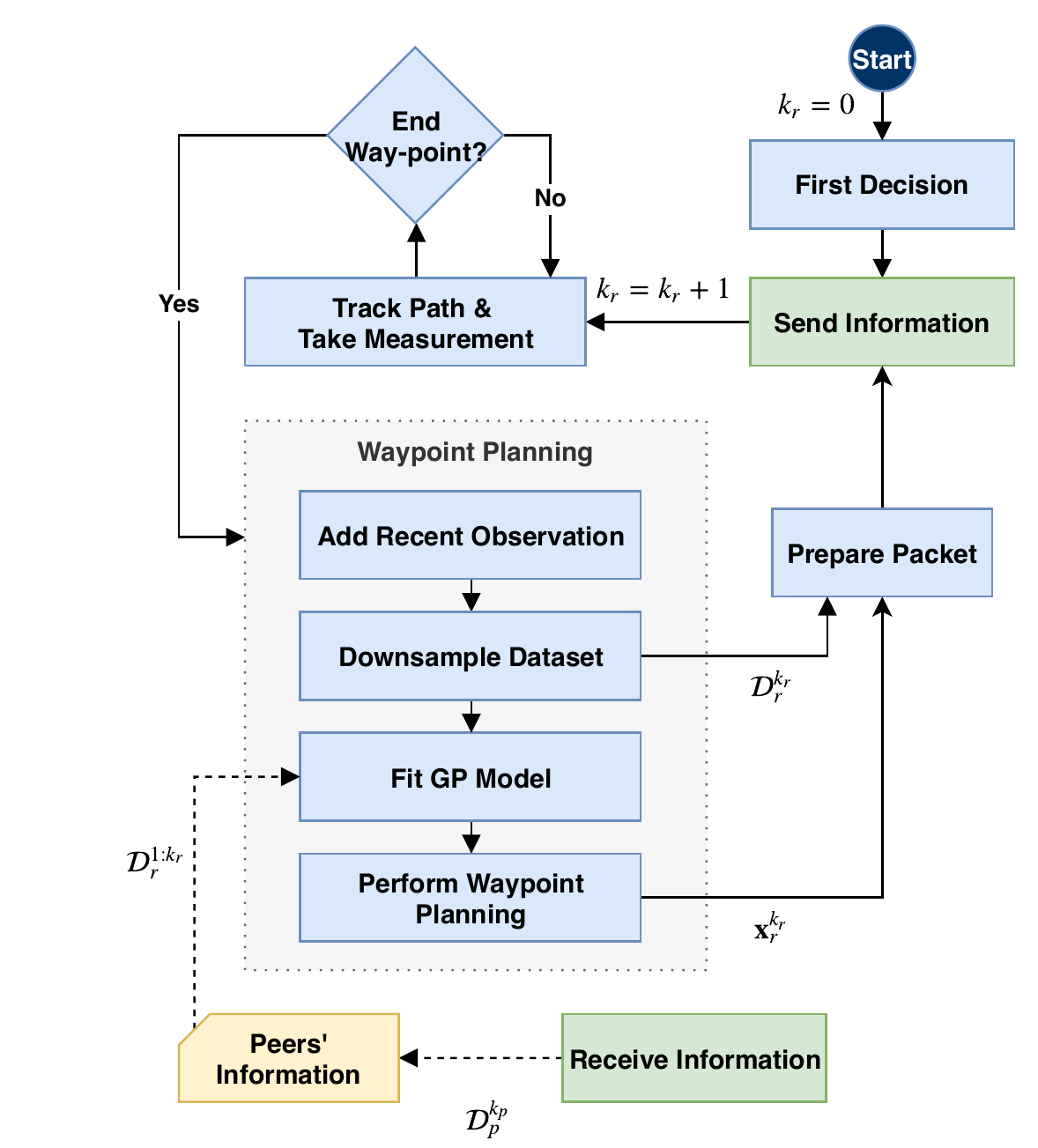}
\caption{\decBayes~architecture for each robot in the swarm}
\label{fig:decBayesflowchart}
\vspace{-0.2cm}
\end{figure}

The robot behaviors including its motion, communication, and decision-making are illustrated in Fig.~\ref{fig:decBayesflowchart} and the pseudocode of our proposed decentralized \decBayes~algorithm is depicted in Alg.~\ref{alg:decBayes}. Each robot in a team of size $N_r$ is assumed to run the \decBayes~algorithm at each decision-making step (i.e., after reaching its waypoint) to take the best action by maximizing an acquisition function that guides the team to the source location over the course of the operation.  Importantly, these decision-making instances need not be synchronized across robots, unlike many existing decentralized implementations.

%
\subsection{Acquisition Function}
Robot-$r$ solves an optimization problem based on its information ($\mathcal{D}^{1:k_r}$ and $\mathbf{\hat{X}}_{-r}^{k_r}$), including self-observations and shared peers' observation from the beginning of the mission till the decision-time $k_r$ ($\mathcal{D}^{1:k_r}=\bigcup_{r=1}^{N_r}\bigcup_{i=1}^{k_r}\mathcal{D}_r^i$; $\mathcal{D}_r^i=[\mathbf{X}_r^i, \mathbf{y}_r^i]$) and the current local peers' next target location ($\mathbf{\hat{X}}_{-r}^{k_r} = \bigcup_{p=1; p\neq r} \hat{X}_{-rp}^{k_r}$). For the $r^\text{th}$ robot, our mathematical formulation of the acquisition function can be expressed as:
\begin{align}
\label{eq:mainObjectiveFunc}
\mathbf{x}^{k+1}_r = \text{arg}\max_{\mathbf{x}\in \mathcal{X}^{k_r}}\Big( \alpha \cdot h_r(\mathbf{x}, \mathcal{D}^{1:k_r}) + (1 - \alpha) g_r(\mathbf{x}, \mathcal{D}^{1:k_r},\mathbf{\hat{X}}_{-r}^{k_r}) \Big)
\end{align}
\vspace{-0.3cm}
s.t.
\vspace{-0.3cm}
\begin{align}
0\leq l_s^{k_r}=\|\mathbf{x} - \mathbf{x}_r^{k_r}\|\leq V T
\end{align}
where the first term, $h_r(.)$, leads robot $r$ to the expected location of the source (exploitation) and the second term, $g_r(.)$, minimizes the knowledge uncertainty of robot $r$. The length ($l_s$) of the path $s$ is bounded based on the decision-horizon $T$ and the nominal velocity of the robots ($V$). The individual terms of the acquisition function are described next.
%
\begin{algorithm}[!b]
\caption{Bayes-Swarm Algorithm}\label{alg:decBayes}
\textbf{Input:} $GP_r, \mathbf{x}_r, X^{k_r}_{-r}$ - the current location and recent observations of the robot ($\mathbf{x}$), and the next waypoints of its peers ($X^{k_r}_{-r}$).\\
\textbf{Output:} $x_r^{k_r+1}$ - the next waypoint of robot-$r$ at its iteration $k_r$.
\begin{algorithmic}[1]
\Procedure{takeDecision}{$r, k_r, N_r, \Delta\theta$}
\If {$k_r = 0$}
\State {$\mathbf{x}_r^{k_r}\gets $ \textsc{takeFirstDecision}($r, k_r, N_r, \Delta\theta$)}
\Else
\If {Size of $\mathcal{D}_r^{k_r} > N_\text{max}$}\Comment{$N_\text{max} = 400$}
\State {Down-sample $\mathcal{D}_r^{k_r}$ to $N_\text{max}$ n observations}
\EndIf
\State {$\mathbf{x}_r^{k_r} \gets $ by solving the optimization, Eq.\eqref{eq:mainObjectiveFunc}}
\EndIf
\State {$k_r \gets k_r + 1$}
\State {\textbf{return} {$\mathbf{x}_r^{k_r}, k_r$}}
\EndProcedure
\Procedure{takeFirstDecision}{$r, N_r, \Delta\theta, V, T$}
\State {$d \gets V T$}
\If {$\Delta\theta = 360$} \Comment{$\Delta\theta$: Initial feasible direction range}
\State {$\theta \gets r\Delta\theta/N_r$}
\Else
\State {$\theta \gets r\Delta\theta/(N_r+1)$}
\EndIf
\State {$\mathbf{x}_r^1 \gets [d \cos\theta, d \sin\theta]$}
\State {\textbf{return} {$\mathbf{x}_r^1$},}
\EndProcedure
\end{algorithmic}
\end{algorithm}
%

\subsection{Source Seeking Term}
The source location is the optimum point of the source signal. In this approach, robots model the source signal using a GP and the location with the maximum expected value based on their then-current GP model of the environment would represent the greedy (exploitive) choice at each waypoint planning instance. Due to the motion constraints of the robot and limited decision-time horizon, all such a location may not be a feasible choices. To consider this factor, we define the source seeking term as follows:
\begin{align}
\label{eq:sourceSeeking}
h_r(\mathbf{x},\mathcal{D}) &= \frac{1}{1+(\mathbf{x} - \mathbf{\bar{x}}^*)^T(\mathbf{x} - \mathbf{\bar{x}}^*)}
\end{align}
\vspace{-0.2cm}
where
\vspace{-0.2cm}
\begin{align}
\mathbf{\bar{x}}^* & = \text{arg}\max_{\mathbf{\bar{x}}}\mu(\mathbf{\bar{x}})
\end{align}

\subsection{Knowledge-gain Term}
As we mentioned in the first section, robots typically gather information over their path; therefore, different paths cause different knowledge-gains. This concept is known as informative path planning (IPP), where robots plan paths such that best/maximum possible information is extracted.
In this paper, we are interested in paths that minimizes the uncertainty in the robots' belief (knowledge), which is analogous to maximizing the knowledge-gain. For this purpose, we are estimating the uncertainty in the belief (modeled by a GP) based on the gathered observations and the planned future observations (other robots' planned paths). We thus define the knowledge-gain as follows:
%
\begin{align}
\label{eq:knowledgeGain}
g_r(\mathbf{x},\mathcal{D},\mathbf{\hat{X}}) &= \int_{s(\mathbf{x})} \sigma(s(u)) du
\end{align}
where, the path is written in the parametric form as:
\begin{align}
s(u) &= u \mathbf{x} + (1-u)\mathbf{x_r^{k_r}};\; u\in[0,1]
\end{align}



%
\begin{table*}[h!]
\centering
\caption{Content, size, and frequency of information shared by robot $r$ via communication across the swarm.} \label{tbl:informationSharing}
\footnotesize
\begin{tabular}{ll}
\toprule[0.12em]
\textbf{Property} & \textbf{Descriptions}\\
\midrule[0.12em]
Inter-robot communication frequency & After each waypoint planning instance\\
\hline
Content of transmitted data & 
~\llap{\textbullet}  Its next location to visit ($\mathbf{x}_r^{k_r}$)\\
& ~\llap{\textbullet} Its observations over the last path ($\mathcal{D}_r^{k_r}$)\\
\hline
Average size of outgoing data packets (with time-horizon 1 min) & 364 Bytes\\
\bottomrule[0.12em]
\end{tabular}
\end{table*}
%

\subsection{Information Sharing}
Inter-robot communication is a key element of any swarm system and robots often require to communicate with each other over an ad-hoc wireless network in outdoor applications. However, given the bandwidth limitations of ad-hoc wireless communication and the energy footprint of wireless communication~\cite{li2008robot}, it is typically desirable to reduce the communication burden. To this end, in our proposed method, the decision-making is allowed to be asynchronous and robots share only a down-sampled set of observations.
Table~\ref{tbl:informationSharing} provides a quick overview of the type and frequency of the information shared by each UAV with all its peers across the swarm. Algorithm~\ref{alg:communication} lists two procedures that each robot uses to share or receive information. Robots then proceed to individually update their respective knowledge model based on their own information and the future plan of its peers. Having presented an overview of the \decBayes~method, the next section introduces its distributed virtual implementation, case studies developed to test the performance of \decBayes, and the corresponding implementation settings that we used.

\begin{algorithm}[!b]
\caption{Communication Procedures}\label{alg:communication}
\begin{algorithmic}[1]
\Procedure{receiveInformation}{$r, p, \mathbf{x}^{k_p}_p, \mathcal{D}_p^{k_p}$}
\State {$\mathcal{D}_r^{1:k_r} \gets \mathcal{D}_r^{1:k_r} \bigcup \; \mathcal{D}_p^{k_p}$}
\State {$\hat{X}_{-rp}^{k_r}(1:2) \gets \hat{X}_{-rp}^{k_r}(3:4)$}
\State {$\hat{X}_{-rp}^{k_r}(3:4) \gets \mathbf{x}^{k_p}_p$}
\State {\textbf{return} {$\mathcal{D}_r^{1:k_r}, \hat{X}_{-rp}^{k_r}$}}
\EndProcedure

\Procedure{sendInformation}{$r, x^{k_r}_r, \mathcal{D}_r^{k_r}$}
\If {$k_r = 0$}
\State {Broadcast $\mathbf{x}^{k_r}_r$} \Comment{4 bytes}
\Else
\State {Broadcast \{$\mathbf{x}^{k_r}_r$; $\mathcal{D}_r^{k_r}$}\} \Comment{$4 + 6T$ bytes} 
\EndIf
\EndProcedure
\end{algorithmic}
\end{algorithm}
%

%
\begin{figure*}[!hpt]
\centering
\subfigure[Case study 1: large arena, convex signal distribution]{\includegraphics[width=0.23\textwidth]{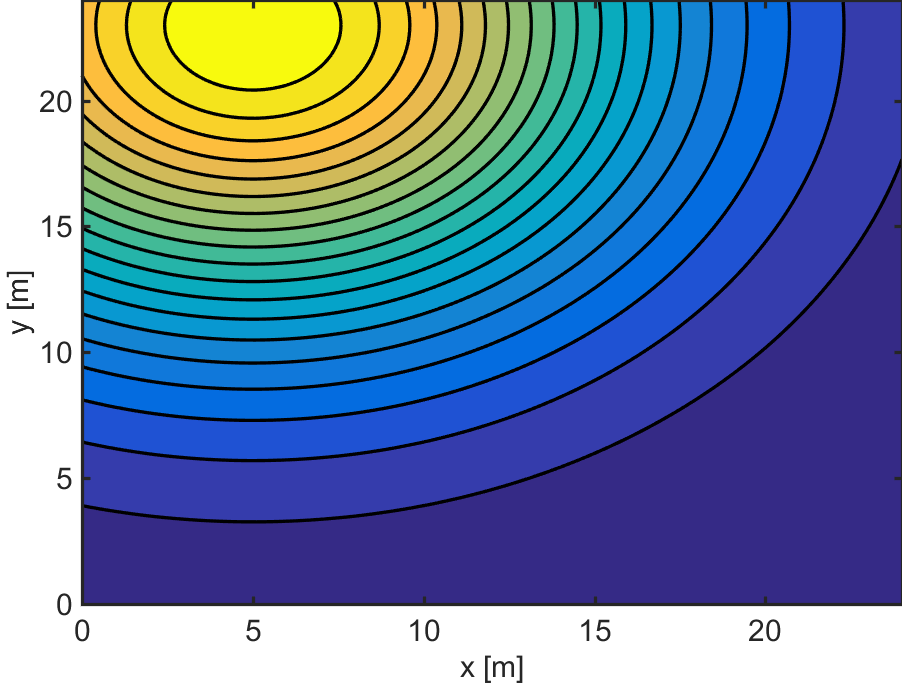}\label{fig:caseStudy1}}\hspace{0.2cm}%
\subfigure[Case study 2: small arena, non-convex signal distribution]{\includegraphics[width=0.23\textwidth]{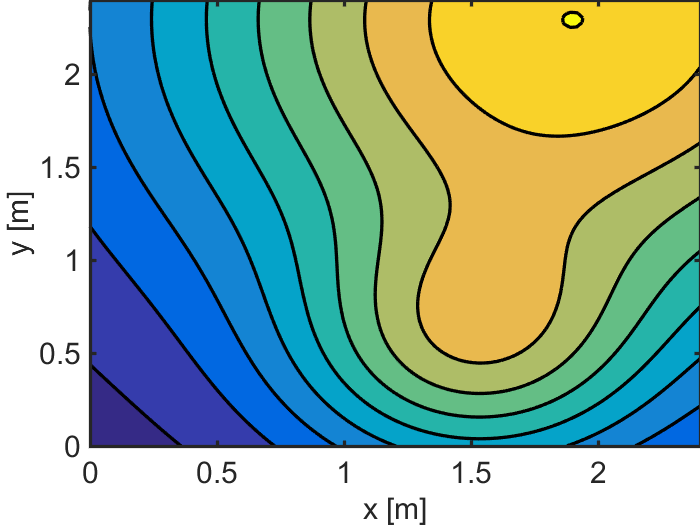}\label{fig:caseStudy2}}\hspace{0.2cm}%
\subfigure[Case study 3: large arena, non-convex signal distribution]{\includegraphics[width=0.23\textwidth]{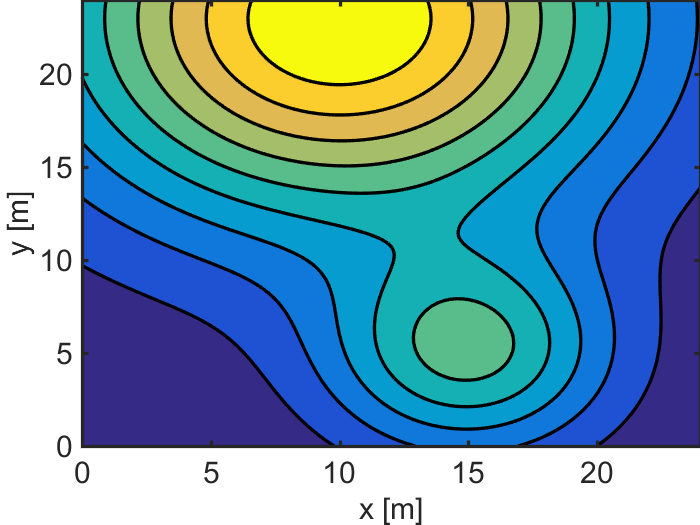}\label{fig:caseStudy3}}\hspace{0.2cm}%
\subfigure[Case study 4: large arena, highly multi-modal signal distribution]{\includegraphics[width=0.23\textwidth]{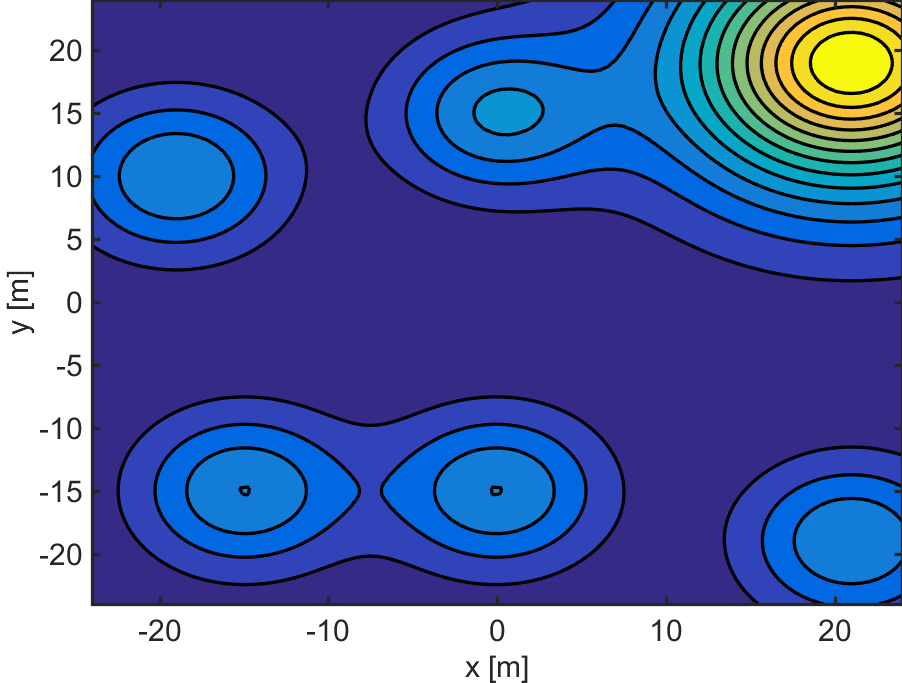}\label{fig:caseStudy4}}
\caption{Four case studies with source distributions of different levels of complexity.}
\label{fig:caseStudies}
\end{figure*}
%
\section{NUMERICAL EXPERIMENTS \& CASE STUDIES}
\subsection{Distributed Virtual Implementation of Bayes-Swarm}
In order to enable a better representation of distributed planning process embodied by a physical swarm of robots, we develop a simulated environment that provisions a parallel computing deployment of \textit{Bayes-Swarm}. This uses \textit{"MATLAB"}'s parallel programming capabilities to invoke 40 dedicated threads. Each robot operates (the behavior illustrated in Fig.~\ref{fig:decBayesflowchart}) in parallel with respect to the rest of the swarm, updating its own knowledge model after each waypoint and deciding its own next waypoint. The entire process is simulated in a virtual environment developed with MATLAB R2017b and is executed on a workstation with Intel\textsuperscript{\tiny\textregistered} Xeon Gold 6148 27.5M Cache 2.40 GHz, 20 cores processor and 196 GB RAM. The simulation time step is set at 1 milliseconds. \textit{Robot settings}: we set the velocity of each swarm robot at 10 cm/s based on the specifications of e-puck 2~\cite{mondada2009puck}. The observation frequency is set at 1 Hz. To keep the computational complexity of refitting the GP low, the size of data ($\mathcal{D}_r^{1:k_r}$) used by each robot is downsampled to 400 (i.e., when it grows beyond 400 in the latter stages of the mission).

\subsection{Case Studies}
We design and execute a set of numerical experiments to investigate the performance of the proposed decentralized \decBayes~approach. In order to provide an insightful understanding of the \decBayes~algorithm, three types of tests are conducted for all case studies and the results are evaluated and compared in terms of completion time, cost incurred by robots, knowledge-gain per robot, and mapping error. Mapping error measures how the estimated response surface using GP deviates from the actual response surface of the source in terms of the Root-Mean Square Error (RMSE) metric. The three tests are described next. {\it{Study 1:}} a parametric analysis to study how the \alphaCoef of \decBayes~affects its performance; {\it{Study 2:}} a scalability analysis is performed to investigate the performance of \decBayes~across multiple swarm sizes; and {\it{Study 3:}} \decBayes~is run using the default values listed in Table~\ref{tbl:caseStudySettings} to analyse its performance regarding different source distributions (i.e., single-modal and multi-modal response surfaces) and results are compared with that of standard \textit{exhaustive search} and \textit{random walk} methods.

Four distinct case studies (Fig.~\ref{fig:caseStudies}) are defined, corresponding to different combinations of source locations, to test the performance and robustness of the \decBayes~method. The first case study is a large convex source signal and the rest of the case studies are non-convex (multi-modal) signal sources. The case study 4 is the most challenging as it contains one global and five local maxima (highly non-convex function).

In this paper, \decBayes~utilizes two termination criteria during operation. The primary criterion terminates the search if any robot arrives within $\epsilon$-vicinity of the source signal location. In addition, \decBayes~terminates if the operation reaches a maximum allowed search time ($T_\text{max}$). The distance threshold $\epsilon$ is set at 5 cm and the maximum search time $T_\text{max}$ is outlined for each case study in Table~\ref{tbl:caseStudySettings}. The decision-time horizon ($T$) is set at 4 seconds for the first decision-making step; then it changes to 10 seconds for the later decision-making steps. 

\begin{table}[bph!]
\centering
\caption{Max. allowed search time, $T_\text{max}$ (in seconds), for case studies.} \label{tbl:caseStudySettings}
\footnotesize
\begin{tabular}{crr}
\toprule[0.12em]
\textbf{Case Study} & \textbf{\decBayes} & \textbf{Random-walk}\\
\midrule[0.12em]
1 & 500 & 4,000\\
2 & 100 & 50,000\\
3 & 500 & 60,000\\
4 & 700 & 60,000\\
\bottomrule[0.12em]
\end{tabular}
\end{table}
%

%
%

%
\begin{figure*}[!hpt]
\centering
\subfigure[Robot 1 at t = 4$^-$s]{\includegraphics[width=0.23\textwidth]{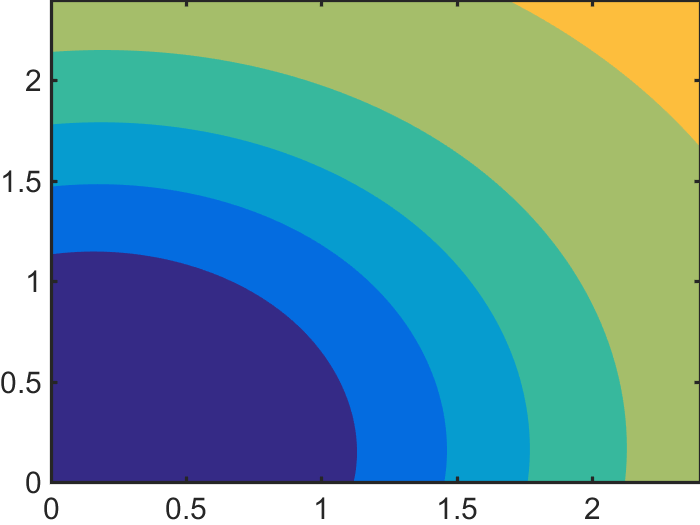}\label{fig:snapshotA}}\hspace{.2cm}%
\subfigure[Robot 4 at t = 4$^+$s]{\includegraphics[width=0.23\textwidth]{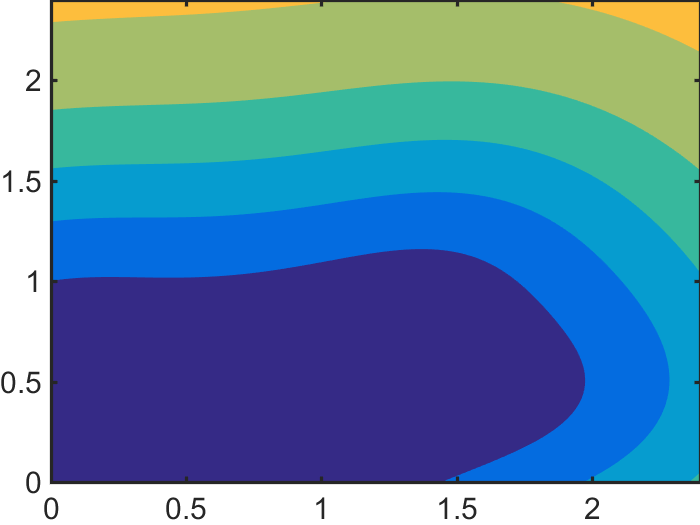}\label{fig:snapshotB}}\hspace{.2cm}%
\subfigure[Robot 3 at t = 26s]{\includegraphics[width=0.23\textwidth]{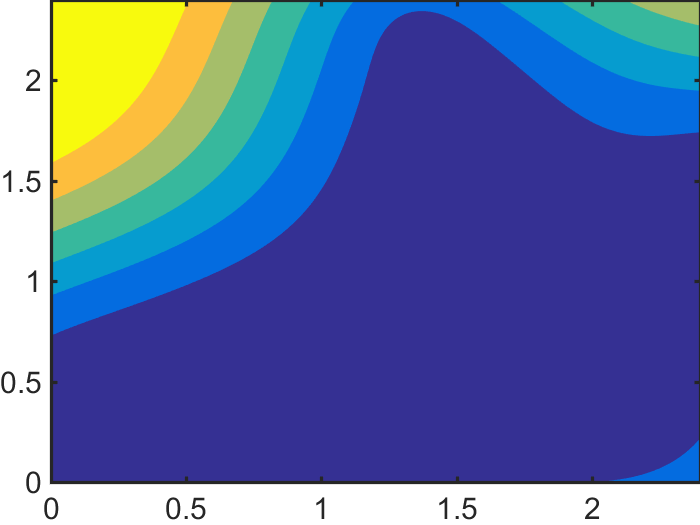}\label{fig:snapshotC}}\hspace{.2cm}%
\subfigure[Robot 3 at t = 54s]{\includegraphics[width=0.23\textwidth]{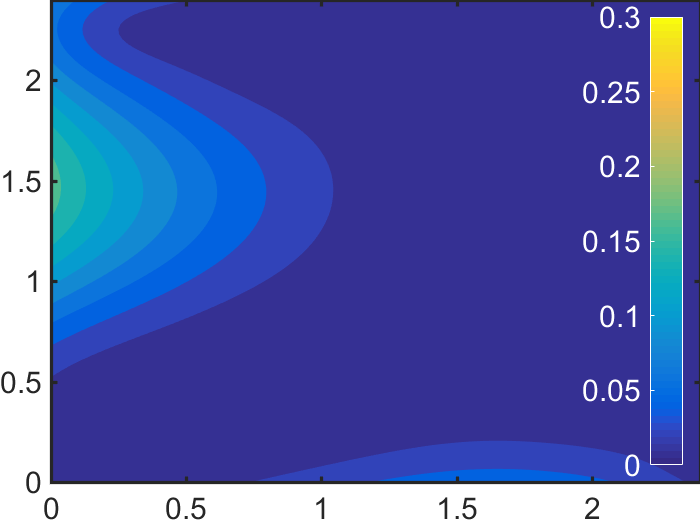}\label{fig:snapshotD}}
\subfigure[Robot 1 at t = 4$^-$s]{\includegraphics[width=0.23\textwidth]{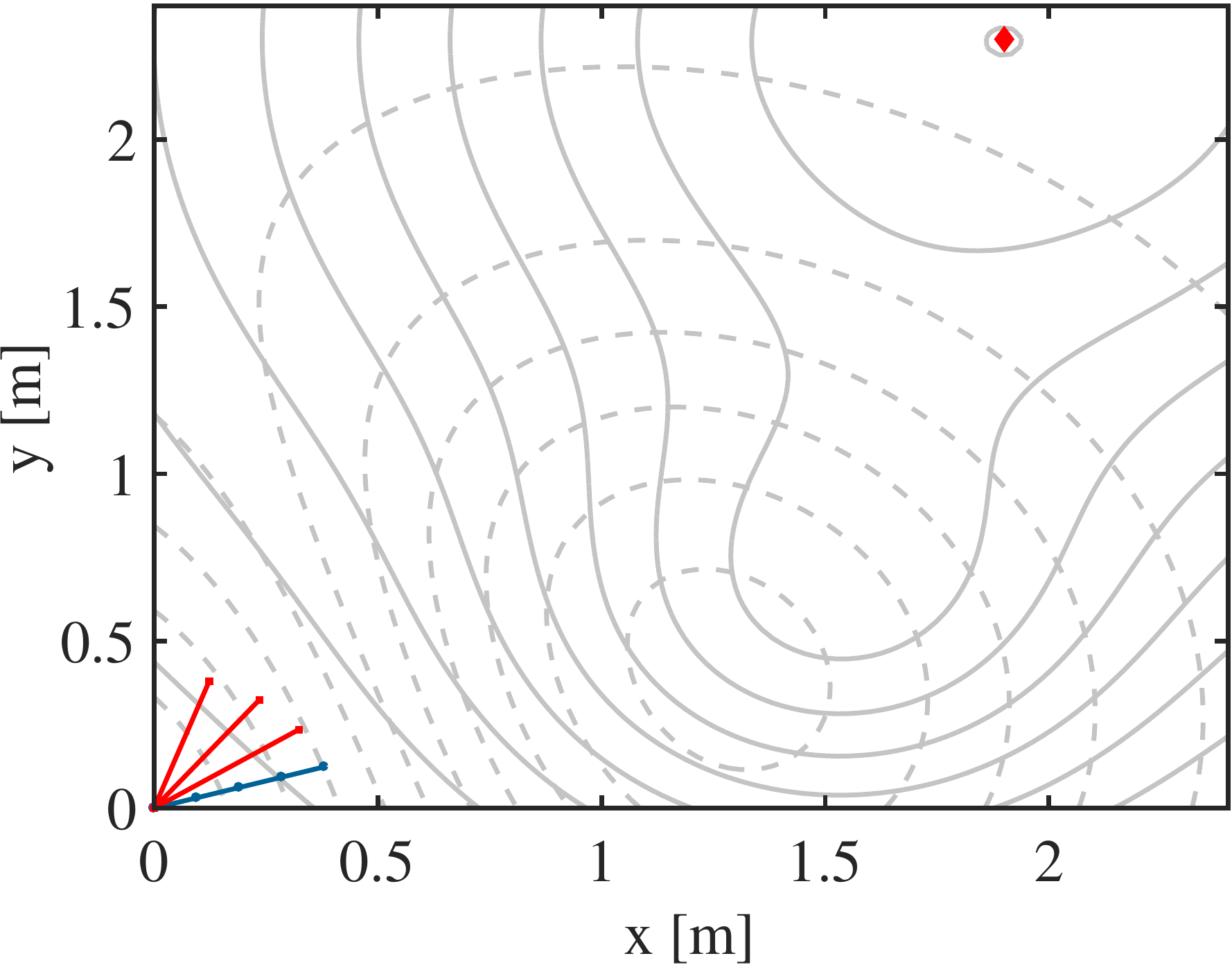}\label{fig:snapshotE}}\hspace{.2cm}%
\subfigure[Robot 4 at t = 4$^+$s]{\includegraphics[width=0.23\textwidth]{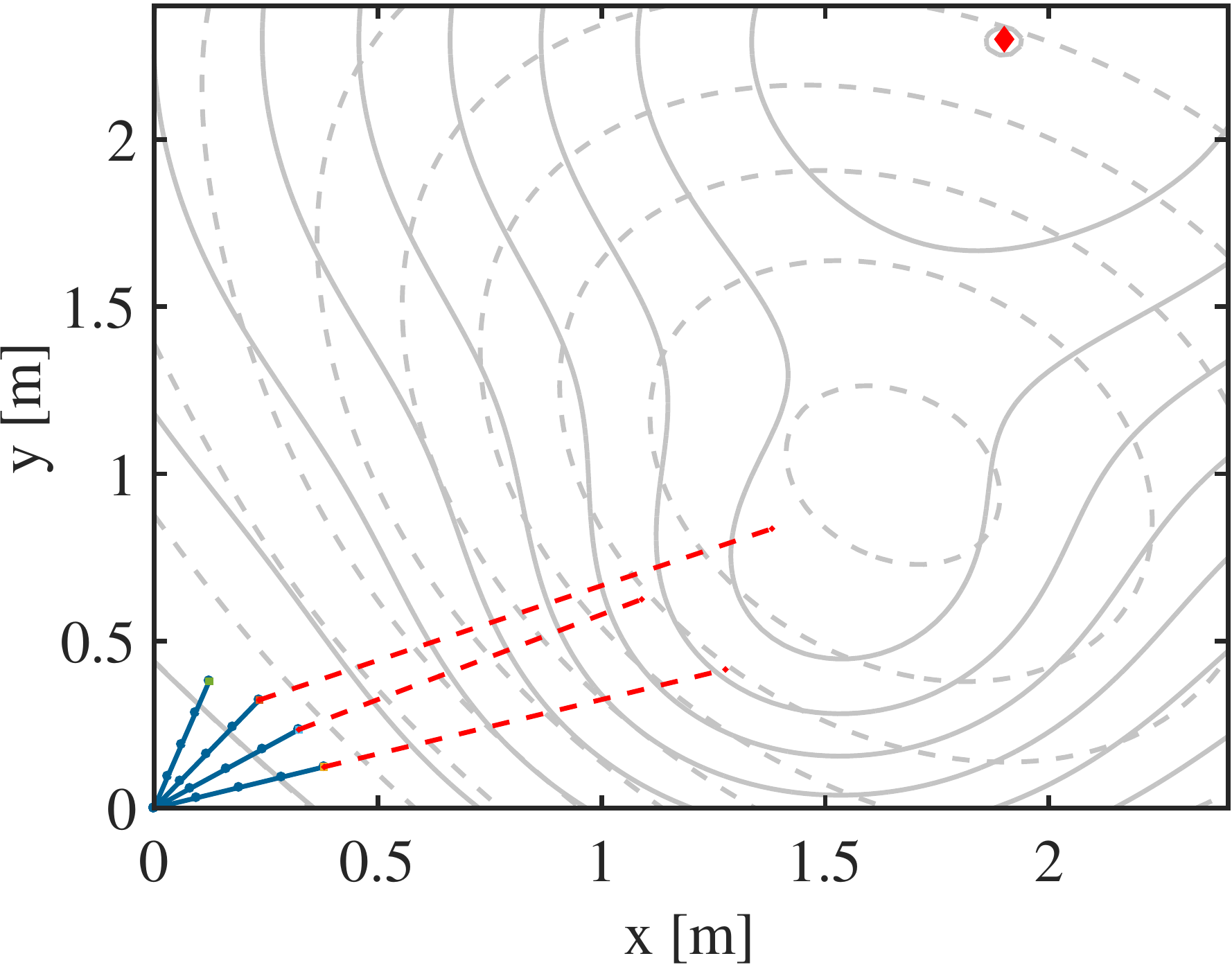}\label{fig:snapshotF}}\hspace{.2cm}%
\subfigure[Robot 3 at t = 26s]{\includegraphics[width=0.23\textwidth]{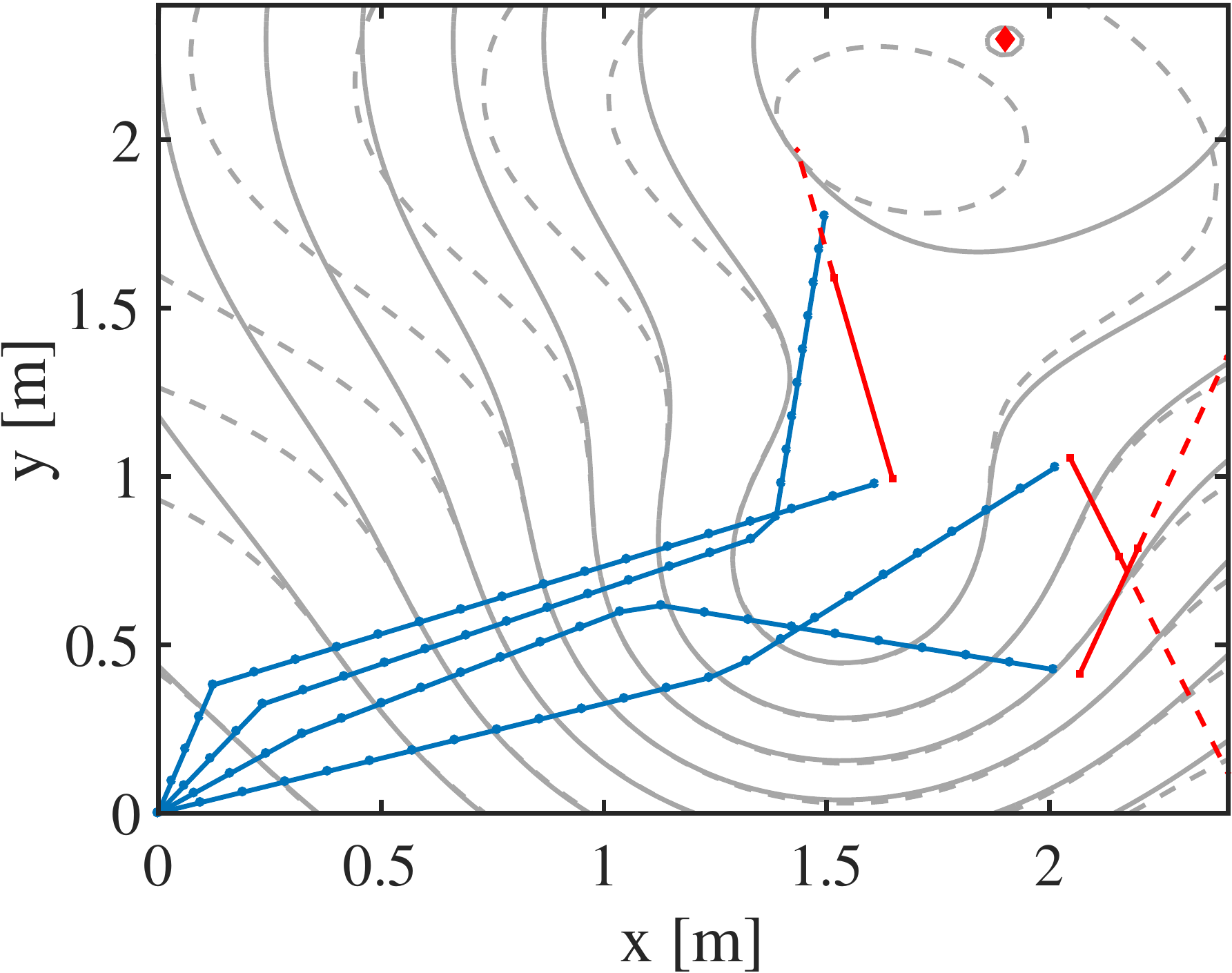}\label{fig:snapshotG}}\hspace{.2cm}%
\subfigure[Robot 3 at t = 54s]{\includegraphics[width=0.23\textwidth]{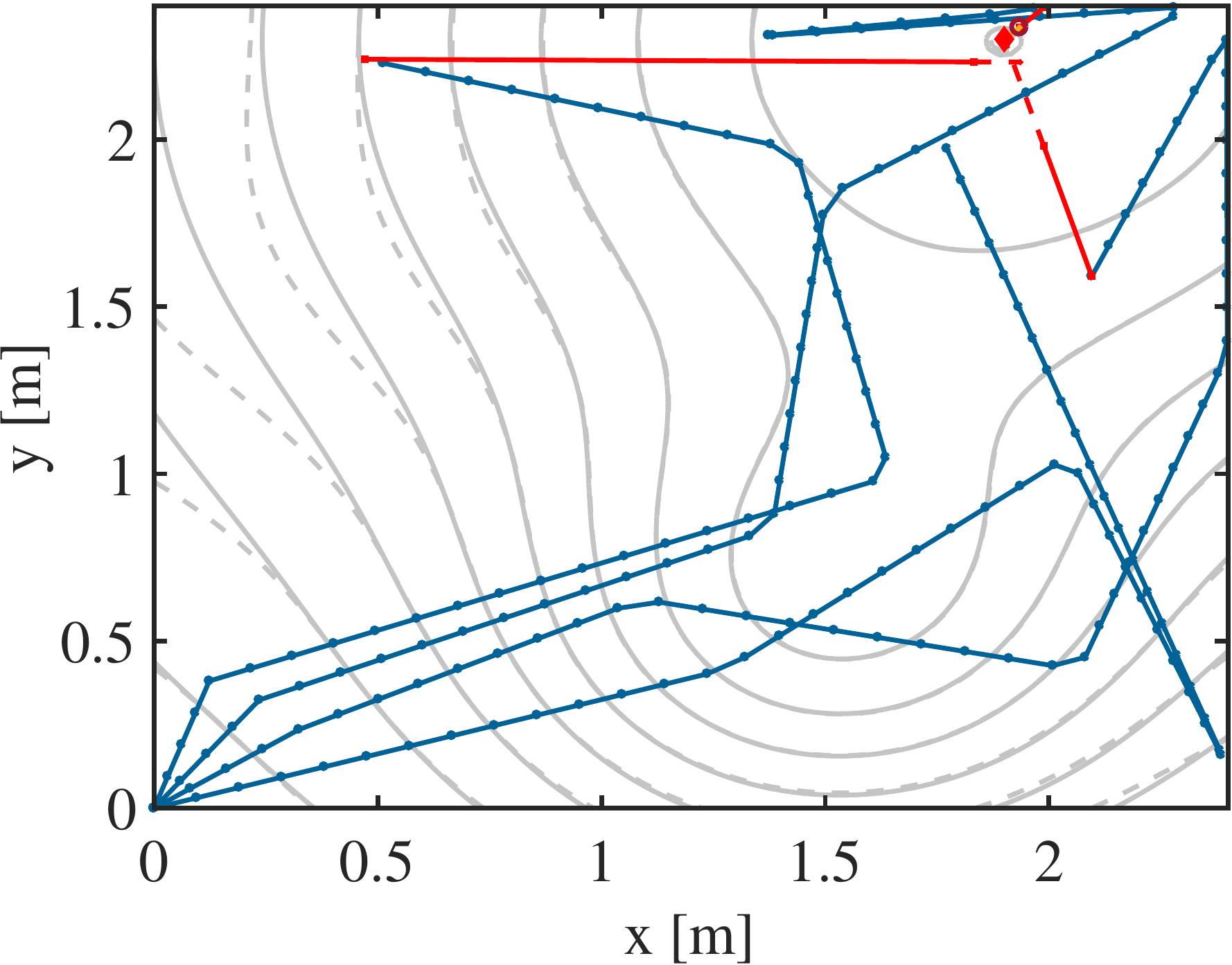}\label{fig:snapshotH}}
\caption{Snapshots for the case study 2 with 4 robots that run \decBayes~with $\alpha = 0.4$. The top figures show the uncertainty map ($\sigma(\mathbf{x})$) and the bottom figures show the robot and knowledge state.
In the bottom figures, the gray solid contours represent the actual source signal (ground truth) and the gray dashed contours represent the source signal (knowledge) model of a robot at the stated time point. Blue solid lines show the paths that robots have already travelled and the observations over which have been shared with all peers, assisting the refitting of their knowledge model. The red solid line shows the paths travelled but the observations over which have not yet been shared with peers. The red dashed lines represent the paths that have been planned but not yet travelled.
}
\label{fig:snapshot}
\end{figure*}
%

\section{RESULTS AND DISCUSSION}\label{sec:results}
\subsection{Overall Performance of Bayes-Swarm}\label{ssec:resultsOverall}
Figure~\ref{fig:snapshot} depicts four snapshots of the \decBayes~for case study 2 with 4 robots and $\alpha = 0.4$. It can be seen from this figure how the estimated knowledge model and its uncertainty improves by exploring the search space. The top figures show the uncertainty map ($\sigma(\mathbf{x})$) and the bottom figures show the robot location and its knowledge state (dashed contours).
In the bottom figures, the gray solid contours represent the actual source signal (ground truth) and the gray dashed contours represent the source signal (knowledge) model of a robot at the stated time point. Blue solid lines show the paths that robots have already travelled and the observations over which have been shared with all peers, assisting the refitting of their knowledge model. The red solid line shows the paths travelled but the observations over which have not yet been shared with peers. The red dashed lines represent the paths that have been planned but not yet travelled. 

From Fig. \ref{fig:snapshotA}-\ref{fig:snapshotE}, it can be seen that when Robot 1 reaches its first waypoint, only 4 self observations are available to it;, hence it is able to build only a relatively inaccurate knowledge model (that gives the expected location of the source at $(1.6,1.0)$, which is in reality far away from both of the actual sources). When the last robot (robot 4) takes decision, it has its peers' observations at $t=4^+s$. The knowledge model (Fig.~\ref{fig:snapshotF}) is still inaccurate, but the uncertainty map (Fig.~\ref{fig:snapshotB}) is improved. After 26 seconds (Figs.~\ref{fig:snapshotC} and \ref{fig:snapshotG}), the robots are able to converge to a fairly accurate knowledge model of the signal distribution, and their future updates and planning (seen in Figs.~\ref{fig:snapshotD} and \ref{fig:snapshotH}) puts two robots in the team within the threshold of the source location at time $t=54s$. 


%
\begin{figure*}[!hpt]
\centering
\subfigure[$\alpha = 0$]{\includegraphics[width=0.35\textwidth]{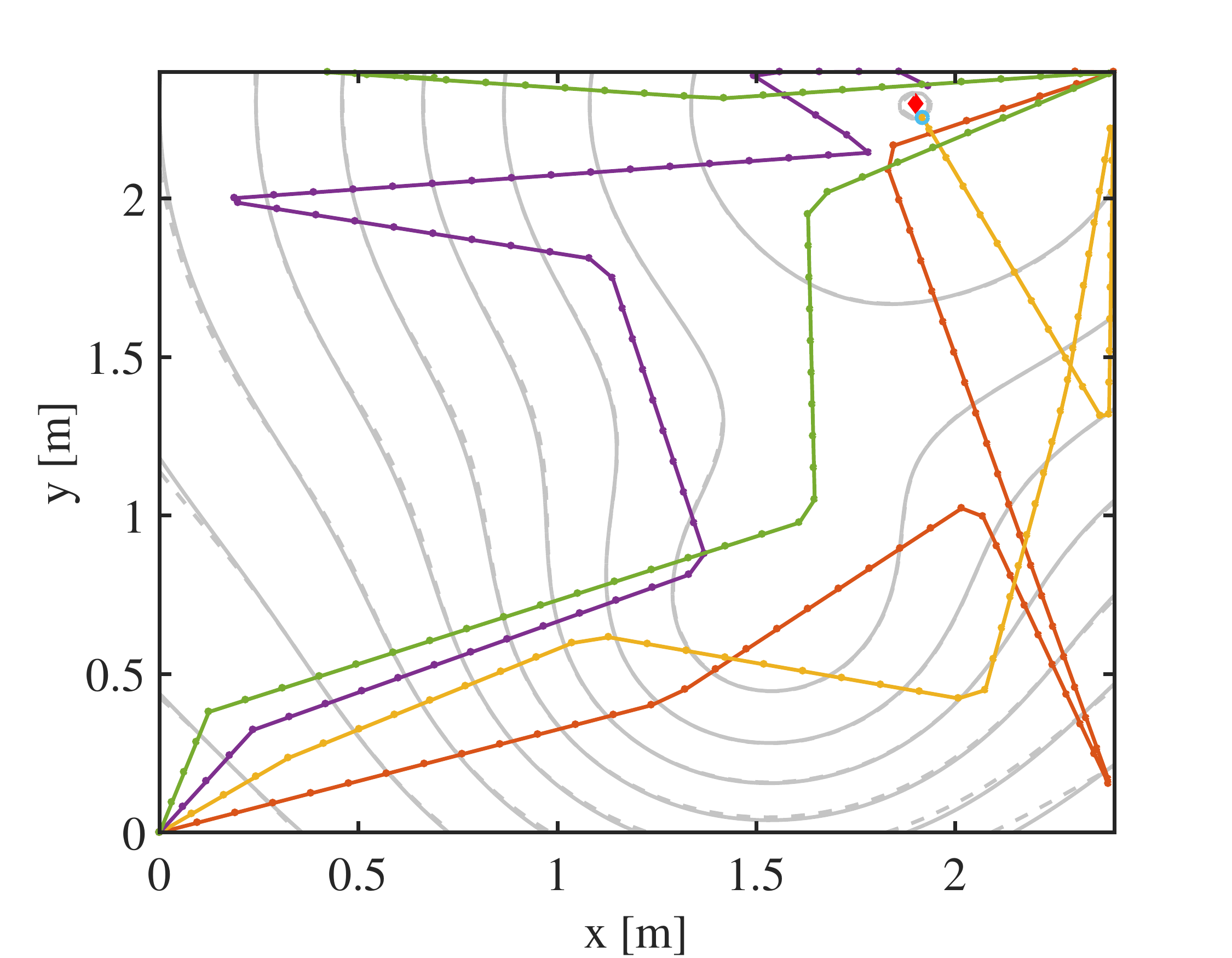}\label{fig:studyAlpha0}}%
\subfigure[$\alpha = 0.4$]{\includegraphics[width=0.35\textwidth]{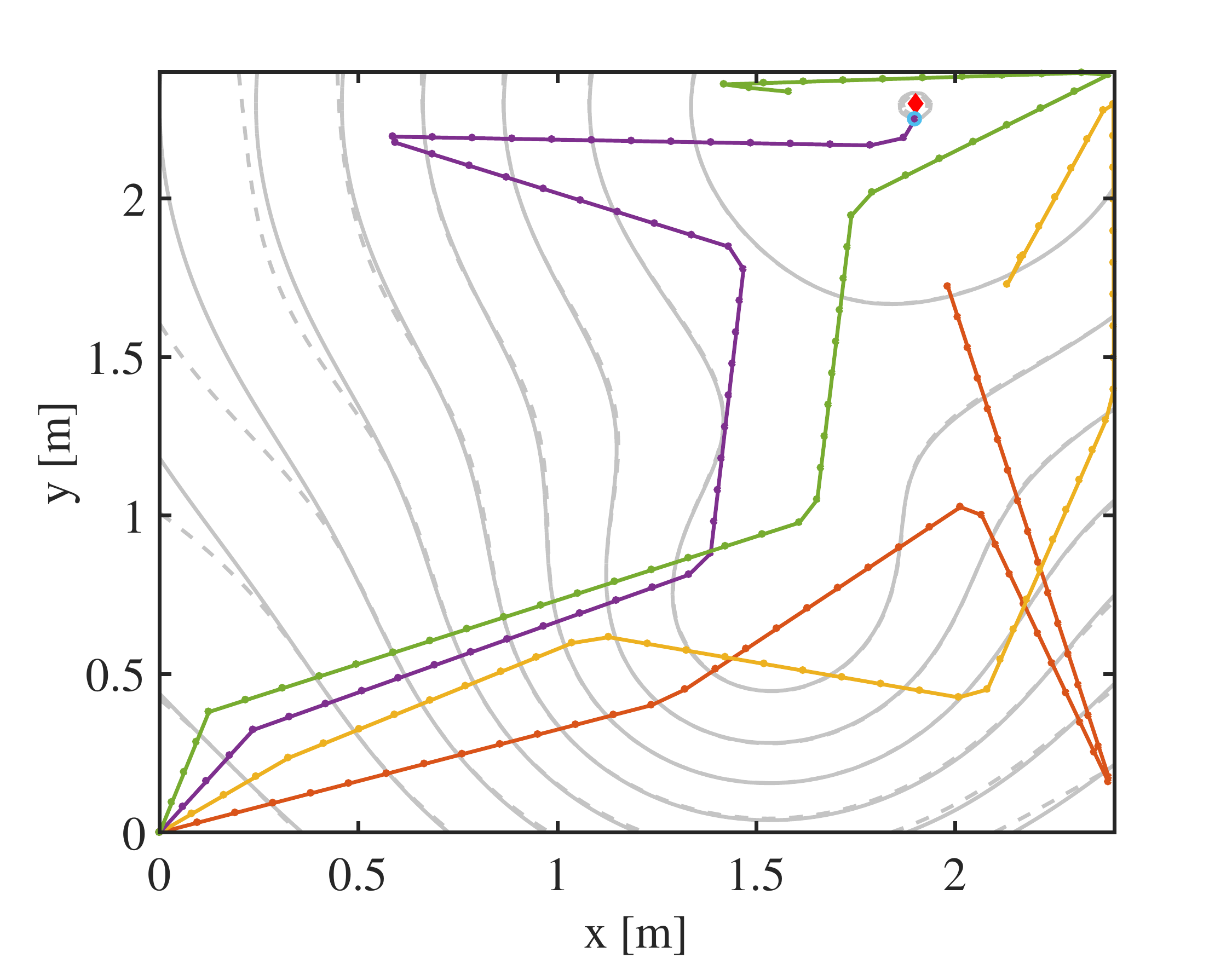}\label{fig:studyAlpha4}}%
\subfigure[$\alpha = 1.0$]{\includegraphics[width=0.35\textwidth]{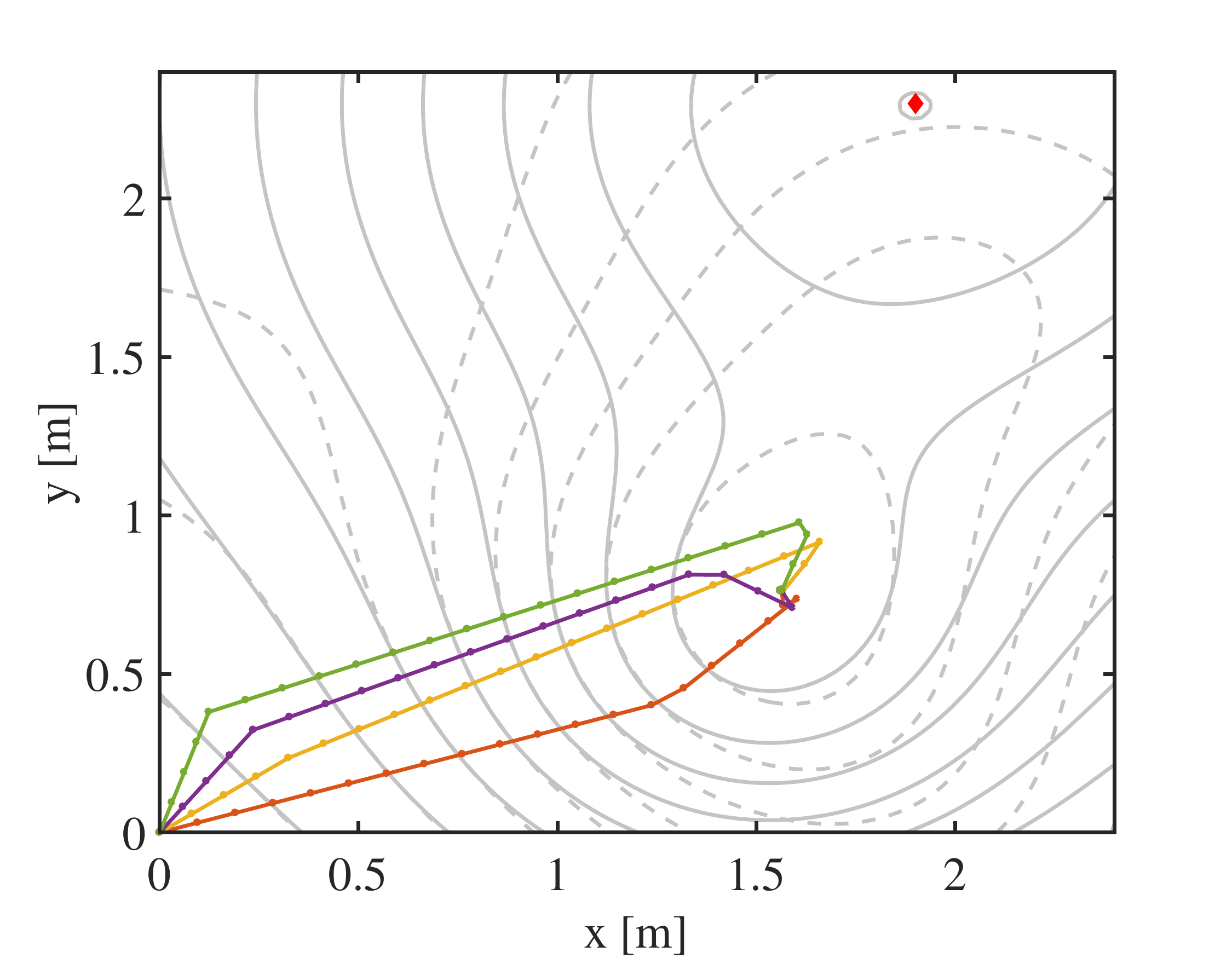}\label{fig:studyAlpha10}}
\caption{Performance dynamics of a 4-robot team under different values of the exploitation/exploration balance coefficient ($\alpha$), for case study 2. The solid line is the actual source signal distribution (ground truth) and the dashed line represents the extracted source signal distribution (knowledge) modeled by the robots. The red dot shows the actual source location.}
\label{fig:studyAlphaShowCase}
\end{figure*}
\begin{figure*}[!hpt]
\centering
\subfigure[Case study 2: completion time]{\includegraphics[width=0.4\textwidth]{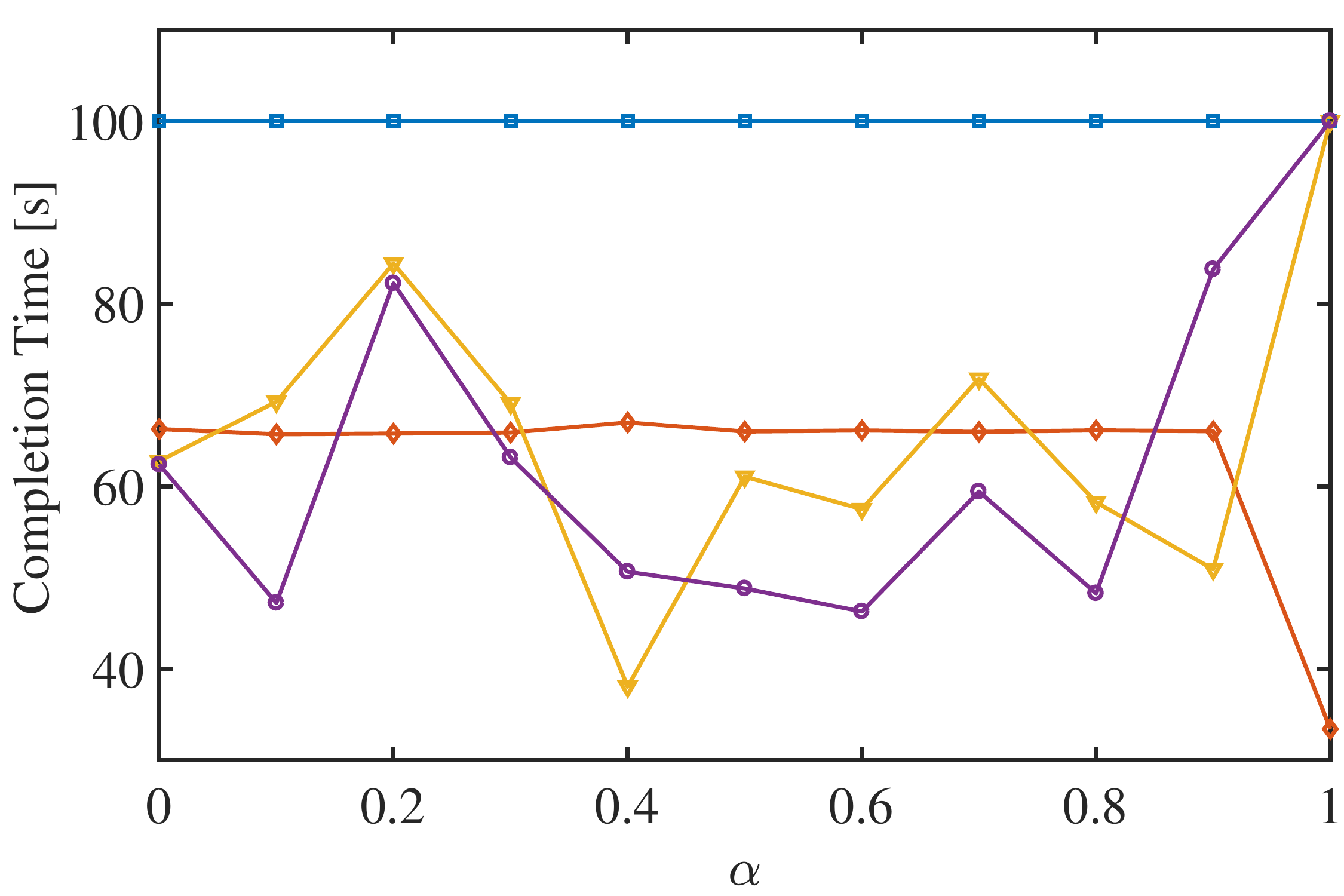}\label{fig:studyAlphaCS2CT}}\hspace{1cm}%
\subfigure[Case study 2: knowledge model accuracy]{\includegraphics[width=0.4\textwidth]{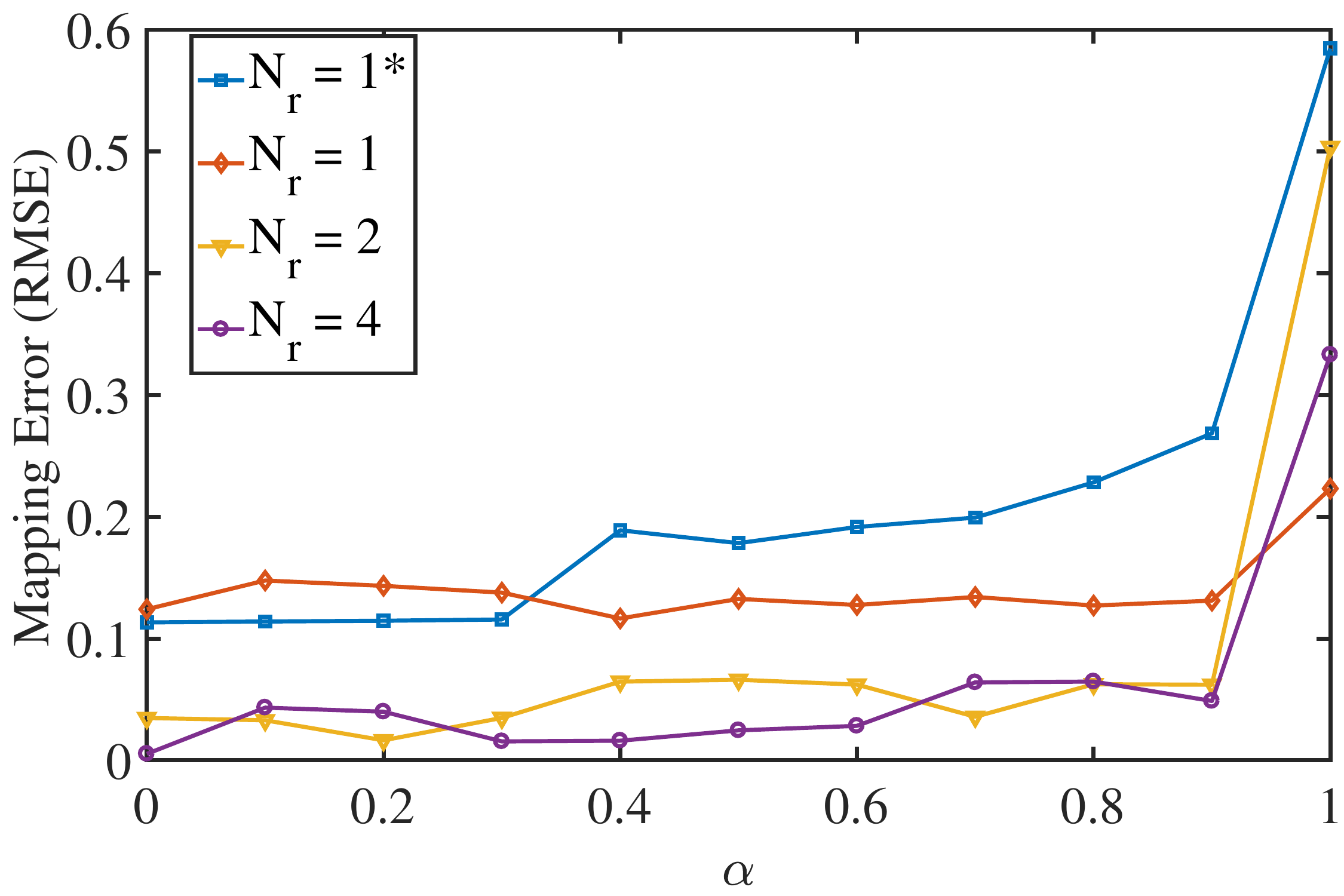}\label{fig:studyAlphaCS2ME}}
\caption{Parametric analysis of the exploitation/exploration balance coefficient ($\alpha$) in case study 2 (small arena, bi-modal signal distribution). For all runs, the maximum allowed search time is set at 100 seconds ($T_\text{max} = 100$). In this case study, we provided two 1-robot scenarios, $N_r=1^*$ and $N_r=1$, by setting the initial feasible direction ($\Delta\theta$) at 0\deg and 45\deg, respectively, to demonstrate the sensitivity of single robot performance on the initial uninformed action.}
\label{fig:studyAlphaCS2}
\end{figure*}
%

\subsection{Study 1: Parametric Analysis of Bayes-Swarm}\label{ssec:resultsParamAnalysis}
\begin{figure*}[!hpt]
\centering
\subfigure[Case study 4: completion time]{\includegraphics[width=0.4\textwidth]{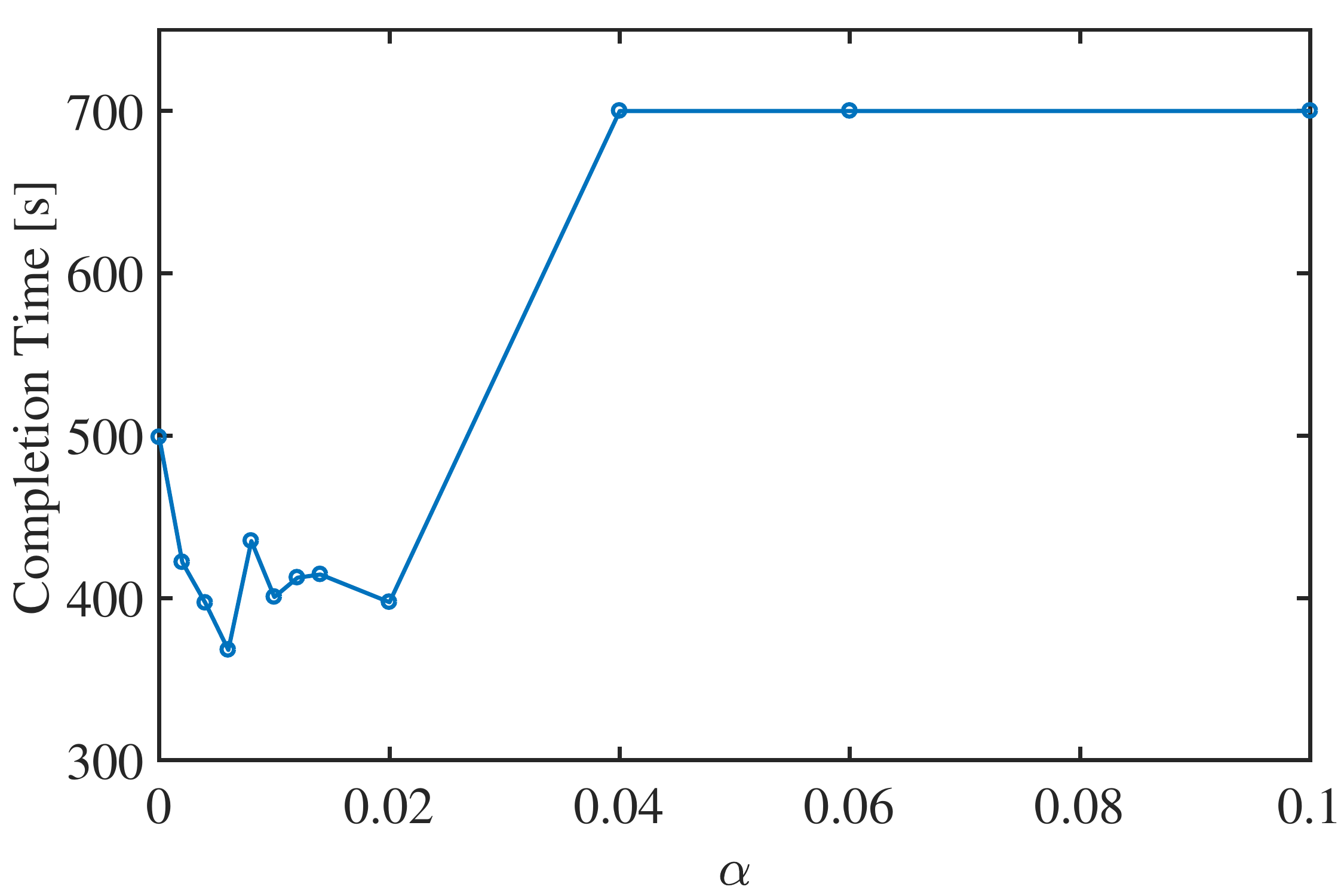}\label{fig:studyAlphaCS4CT}}\hspace{1cm}%
\subfigure[Case study 4: knowledge model accuracy]{\includegraphics[width=0.4\textwidth]{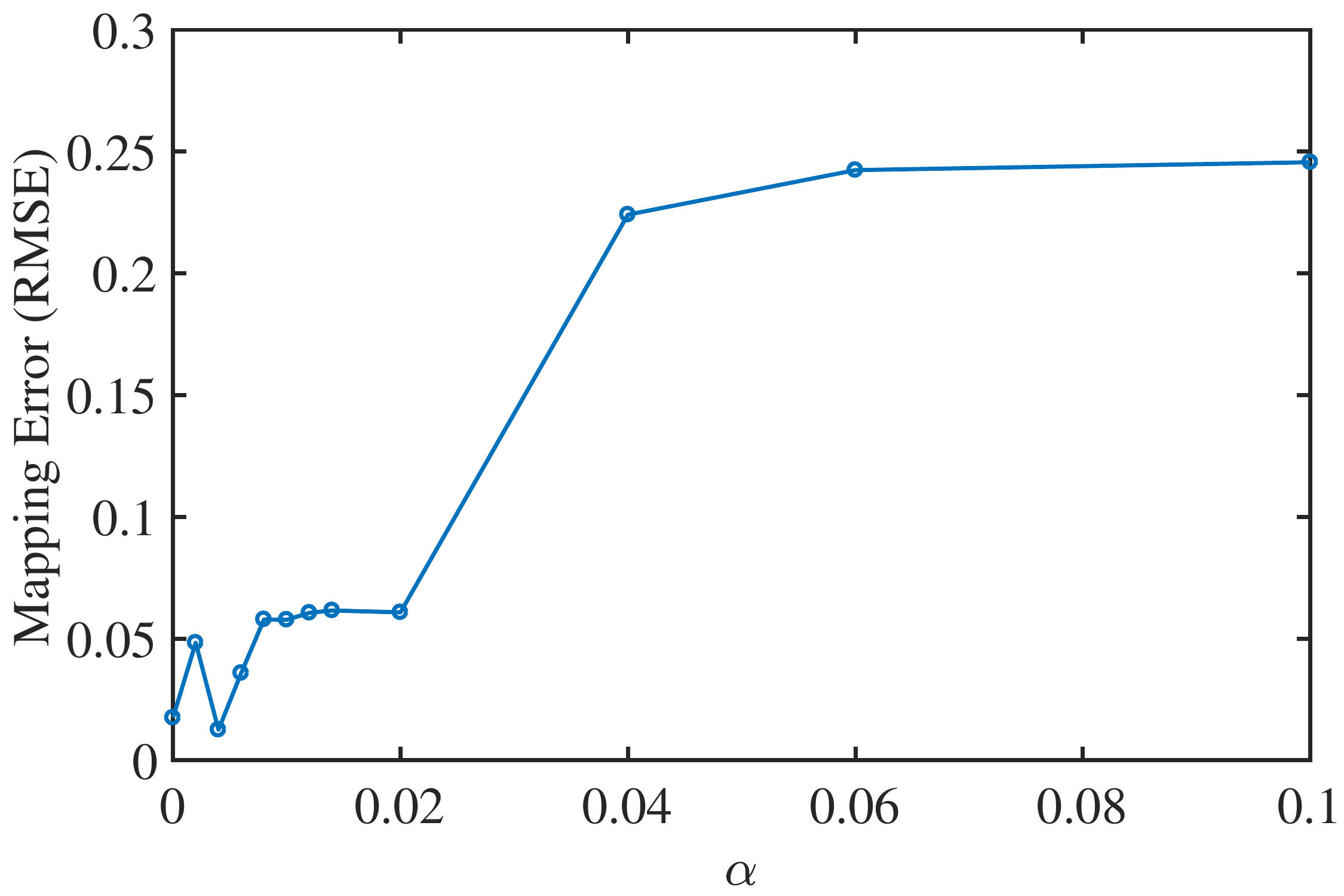}\label{fig:studyAlphaCS4ME}}
\caption{Performance dynamics of a 10-robot team under different values of the exploitation/exploration balance coefficient ($\alpha$), case study 4 (large arena, highly multi-modal signal distribution). For all runs, the maximum allowed search time is set at 700 seconds ($T_\text{max} = 700$).}
\label{fig:studyAlphaCS4}
\end{figure*}

In the proposed decentralized method, there is one major prescribed parameter that needs to be prescribed or tuned -- the \alphaCoef parameter $\alpha$, that regulates the balance between exploration and exploitation. We run an experiment to study how this \alphaCoef parameter ($\alpha$ varying from 0 to 1) affects the performance of \decBayes~for the case studies 2 and 4, across multiple swarm sizes. Snapshots of the final state of robots for three values of $\alpha$ for the case study 2 with 4 robots are depicted in Fig.~\ref{fig:studyAlphaShowCase}. The performance outcomes in terms of completion time, and mapping error are summarized in Figs.~\ref{fig:studyAlphaCS2}-\ref{fig:studyAlphaCS4}.

\textit{\textbf{Pure source seeking ($\alpha = 1$):}} One of the extreme case happens when the knowledge-gain term is eliminated in the objective function; in this mode, robots try to reach the expected source location faster without exploring the area (getting enough knowledge) - basically the purely greedy approach. For this purpose, the \alphaCoef is set at $\alpha=1$. Figure~\ref{fig:studyAlpha10} illustrates the behavior of robots under this setting. It can be seen from this figure that, the estimated source signal or knowledge model is quite inaccurate, due to the lacking of explorative search.

\textit{\textbf{Only knowledge-gain term ($\alpha = 0$):}} By setting $\alpha=0$, the objective function (Eq.~\eqref{eq:mainObjectiveFunc}) is reduced to the knowledge-gain term (Eq.~\eqref{eq:knowledgeGain}), which results in purely explorative search. As expected, under this setting robots are ab;e to estimate a relatively accurate model of signal distribution (Fig.~\ref{fig:studyAlpha0}). This mode is suited for mapping applications, such as mapping offshore oil spills \cite{odonkor2019distributed}.

\textit{\textbf{Combined source seeking \& knowledge-gain terms -- different trade-offs ($0 < \alpha < 1$):}} By setting the \alphaCoef $\alpha$ at values between 0 and 1, we can tune the degree of exploration and exploitation of the swarm search. Figures~\ref{fig:studyAlphaCS2ME}-\ref{fig:studyAlphaCS4ME} show that, by increasing the \alphaCoef from 0 to 1, the mapping error increases, especially for $\alpha$ values beyond 0.3. Figure~\ref{fig:studyAlpha4} depicts the search behavior of the swarm for $\alpha=0.4$. In this setting, one robot successfully reaches the source location while other robots are still exploring the search area. Depending on the complexity of the source signal distribution, the effect of \alphaCoef parameter on the estimation of the knowledge model will vary. 

In terms of completion time, the complexity of the source signal distribution and the initial path of robots play an important role. 
In case study 2, the impact of $\alpha$ on completion time varies with the size of the robot team (Fig. \ref{fig:studyAlphaCS2CT}). 
In case study 4, we can see from Fig.~\ref{fig:studyAlphaCS4CT} that Bayes-Swarm with $\alpha>0.04$ is not able to lead the robots to find the target/source within the maximum allowed time (700 seconds). In order to get the best performance, the \alphaCoef ($\alpha$) needs to be less than 0.02. This is attributed to the need for greater exploration in a multimodal environment.
In summary, for choosing the correct value of $\alpha$ to get the best performance, we need to consider the number of robots, the complexity of the source signal distributions, and the robots' capabilities.

\subsection{Study 2: Scalability Analysis of Bayes-Swarm}\label{ssec:resultsScalabAnalysis}
\begin{figure*}[!hpt]
\centering
\includegraphics[width=1.0\textwidth]{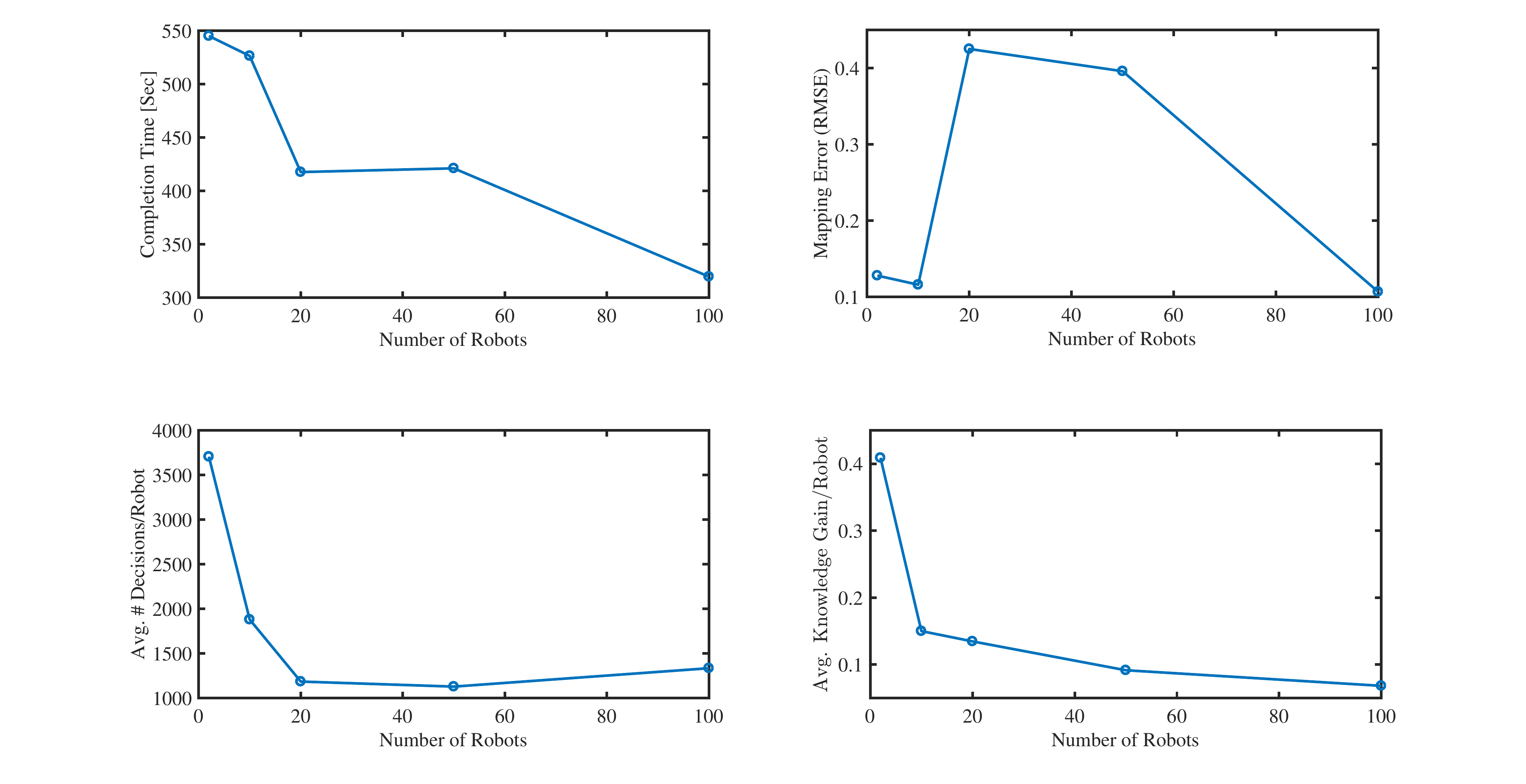}
\caption{Scalability analysis of \decBayes~with $\alpha = 0.4$ and swarm sizes varying from 2 to 100, when applied to case study 4 environment.}
\label{fig:studying_nrobot}
\vspace{-0.6cm}
\end{figure*}

In this test, we use case study 4 to perform an analysis of how the size of the robot swarm impacts \decBayes's performance. To this end, we run \decBayes~simulations with $\alpha = 0.4$ and swarm sizes varying from 2 to 100. Figure~\ref{fig:studying_nrobot} illustrates the results of this analysis in terms of the completion time, averaged knowledge-gain of each robot ($\bar{g}(x)$), averaged number of decisions per robot ($\bar{N}_d$) and mapping error. The results show that the performance improves by increasing the size of the swarm from 2 to 100, with completion time reducing by $\sim 41.3$\%. Moreover, the averaged number of decisions (waypoint planning instances) per robot and the averaged knowledge-gain per robot respectively decrease by about $64$\% and $83.3$\% when the swarm size grows from 2 to 100.
Although the mapping error with 100 robots is $16.6$\% less than the mapping error with 2 robots, increasing the number of robots does not universally improve the mapping error, as evident from the non-monotonic trend seen in the top right plot of Fig. \ref{fig:studying_nrobot} (unless $\alpha$ is tuned based on the size of swarm).

To summarize the observations made from Fig. \ref{fig:studying_nrobot}, increasing the size of swarms become increasingly effective for complex signal distribution environments. However, beyond a certain swarm size ($\sim$20 in this analysis), there is a decreasing rate of improvement. These observations provide strong evidence of the scalability of the \decBayes~method. At the same time, they highlight the importance of identifying suitable team sizes for suitable mission profiles, given resource constraints and time sensitivity of the mission.

%


%
\begin{table}[t!]
\centering
\caption{Performance of the \decBayes, baseline, and competing algorithms on four test case studies with 5 robots; the ~\alphaCoef of the \decBayes~is set at 0.4 for all case studies.}\label{tbl:compareMethods}
\vspace{.2cm}
\footnotesize
\begin{tabular}{clrr}
\toprule[0.12em]
\textbf{Case Study} & \textbf{Algorithm} & \textbf{Total Time $^*$ [s]} & \textbf{Success Rate}\\
\midrule[0.12em]
\multirow{3}{*}{1} & \decBayes & 246.1 & 1/1\\
& Random-Walk & 20,394 & 1/5\\
& Exhaustive Search & 22,174 & 1/1\\
\midrule[0.12em]
\multirow{3}{*}{2} & \decBayes & 42.5 & 1/1\\
& Random-Walk & 227.6 & 5/5\\
& Exhaustive Search & 225.3 & 1/1\\
\midrule[0.12em]
\multirow{3}{*}{3} & \decBayes & 260.1& 1/1\\
& Random-Walk & - & 0/5\\
& Exhaustive Search & 22,174 & 1/1\\
\midrule[0.12em]
\multirow{3}{*}{4} & \decBayes & 373.2 & 1/1\\
& Random-Walk & - & 0/5\\
& Exhaustive Search & 9,163 $^\dagger$ & 1/1\\
\bottomrule[0.12em]
\end{tabular}
\begin{flushleft}
$^*$ As all random-walk runs are not able to find the source, we only report the total time of the best solution obtained using the random-walk.\\
$^\dagger$ For this case, we divide the search space into four equal quarters and each robot does an exhaustive search in each portion.
\end{flushleft}
\vspace{-0.7cm}
\end{table}

\subsection{Study 3: Comparative Analysis of Bayes-Swarm}\label{ssec:resultsComparative}
As mentioned before, \textit{exhaustive search} and \textit{random-walk} algorithms are implemented beside the \decBayes~for comparative analysis. We test these algorithms to find the source location in the four case studies, illustrated in Fig.~\ref{fig:caseStudies}. The settings of the \decBayes~are not not individually tuned for each case to allow fair comparison; the \alphaCoef is set at 0.4 and $T$ at 4 seconds. Table~\ref{tbl:compareMethods} summarizes the results of this study in terms of the completion time. In this study, the maximum allowed search time of the random-walk search is adjusted to 1.5 times of what is needed by exhaustive search for each case study environment. In case study 4, we partition the arena into 4 parts and each robot searches one part using the exhaustive search method. Note that, in this table, we only report the best performance across 5 runs of the random-walk method for each case. 

The results show that the \decBayes~performs significantly better than exhaustive search and random-walk approaches in all the four case studies. Due to complexity of some of the search environments, the random-walk method often fails to find the source location within the allowed maximum search time, as evident from its poor success rate in Cases 1, 3 and 4. 
The table shows that \decBayes~finds the primary source location about 5 to 100 times faster than the exhaustive search in all four cases. As the random-walk reaches the goal only in the first two case studies, we compare \decBayes~with the random-walk method only in these case studies; \decBayes~is observed to perform 83 and 5 times faster than the random-walk method in case studies 1 and 2.

\section{CONCLUSION}
In this paper, we proposed an asynchronous, decentralized algorithm to perform searching for the source of a spatially distributed signal in 2D arenas, using robot swarms. 
This algorithm is founded on an extension of the batch Bayesian search method, with advancements made for application to embodied swarm systems. A new acquisition function is designed to be able to uniquely incorporate the following: 1) modeling knowledge gain over trajectories, as opposed to at points; 2) implicitly mitigating overlapping trajectories among robots to maximize unique knowledge gain; 3) incentivising robots to reach (closest to) the expectation of the source, while accounting for constraints on the robot's motion and cost incurred by it in reaching a candidate waypoint. A heuristic (weight coefficient, $\alpha$) is currently used to balance the source seeking and knowledge gain components of the acquisition function, and thus further parametric analyses is performed to understand the impact of this coefficient. It is found that suitable values of this parameter depends both on the size of the swarm and the complexity of the signal spatial distribution. An important direction of future research will be to build on this understanding to formulate a situation-adaptive variation (instead of user prescription) of the weighting coefficient. 

To evaluate and compare the performance of the proposed algorithm, {\decBayes}, exhaustive search and random-walk baselines are considered. These algorithms are tested on four distinct case studies, with varying arena size and complexity (non-convexity) of the spatial distribution of the signal. Performance is analyzed in terms of completion time and mapping error. \decBayes~easily outperforms the exhaustive search and random-walk approaches by achieving up to 90 times better values of completion time. 
Scalability of the \decBayes~is also analyzed, with significant performance gain (in terms of superlinear reduction in completion time) observed as the swarm size is changed from 2 to 20, and then mostly saturating owing to the bounds on the size of the arena. It is important to note that increased swarm size (while beneficial to the mission) also increases the rate at which signal data is collected, thus increasing the online computational cost of updating the GP by every robot. Thus future work will also look at approximate (downsampling-based) update approaches, in the cases of applications where 100's to 1000's of robots are needed, or where longer mission time periods are needed. This, along with the consideration of partial observability due to communication constraints and physical demonstration, will allow us to more comprehensively explore the scalability of the \decBayes~algorithm.  
\bibliographystyle{asmems4}
\bibliography{payam2018mas}

\end{document}